\documentclass[preprintnumbers, floatfix, preprintnumbers, letterpaper, superscriptaddress,nofootinbib]{revtex4-2}
\pdfoutput=1
\usepackage{graphicx}
\usepackage{microtype}
\usepackage{amsmath}
\usepackage{amssymb}
\usepackage{subfigure}
\usepackage{hyperref}
\usepackage{url}
\usepackage{xcolor}
\usepackage{color}
\usepackage{mathrsfs}
\usepackage{calrsfs}
\usepackage{amsfonts}
\usepackage{latexsym}
\usepackage{ragged2e}
\usepackage{epstopdf}
\usepackage{textcomp}
\usepackage{phaistos}
\usepackage[utf8]{inputenc}
\usepackage{ulem}
\usepackage{float}
\makeatletter
\renewcommand\@makefnmark{\hbox{\@textsuperscript{\normalfont\color{purple}\@thefnmark}}}
\renewcommand\@makefntext[1]{%
  \parindent 1em\noindent
            \hb@xt@1.8em{%
                \hss\@textsuperscript{\normalfont\@thefnmark}}#1}
\makeatother

\usepackage{caption}
\DeclareCaptionJustification{justified}{\leftskip=0pt \rightskip=0pt \parfillskip=0pt plus 1fil}
\definecolor{vividviolet}{rgb}{0.62, 0.0, 1.0}
\definecolor{amaranth}{rgb}{0.9, 0.17, 0.31}
\definecolor{palatinateblue}{rgb}{0.15, 0.23, 0.89}
\definecolor{brightpink}{rgb}{1.0, 0.0, 0.5}
\definecolor{cornflowerblue}{rgb}{0.39, 0.58, 0.93}
\definecolor{deepcarminepink}{rgb}{0.94, 0.19, 0.22}
\definecolor{radicalred}{rgb}{1.0, 0.21, 0.37}

\hypersetup{ linktoc=all,
    colorlinks, linkcolor={palatinateblue},
    citecolor={brightpink}, urlcolor={amaranth}
}

\graphicspath{{Images/}}

\def\sideremark#1{\ifvmode\leavevmode\fi\vadjust{\vbox to0pt{\vss% the remark
 \hbox to 0pt{\hskip\hsize\hskip1em%                          will appear only
 \vbox{\hsize1.5cm\tiny\raggedright\pretolerance10000%          on the side
 \noindent #1\hfill}\hss}\vbox to8pt{\vfil}\vss}}}%
\begin{document}

\title{Geodesic Motion of Test Particles around the Scalar Hairy Black Holes with Asymmetric Vacua}

\author{Hongyu \surname{Chen}}
\email{hongyuchen0306@qq.com}
\affiliation{School of Science, Jiangsu University of Science and Technology, 212100, Zhenjiang, Jiangsu Province, China}

\author{Wei \surname{Fan}}
\email{fanwei@just.edu.cn}
\affiliation{School of Science, Jiangsu University of Science and Technology, 212100, Zhenjiang, Jiangsu Province, China}

\author{Xiao Yan \surname{Chew}}
\email{xiao.yan.chew@just.edu.cn}
\affiliation{School of Science, Jiangsu University of Science and Technology, 212100, Zhenjiang, Jiangsu Province, China}

\begin{abstract}
An asymptotically flat hairy black hole (HBH) can exhibit distinct characteristics when compared to the Schwarzschild black hole, due to the evasion of no-hair theorem by minimally coupling the Einstein gravity with a scalar potential which possesses asymmetric vacua, i.e, a false vacuum $(\phi=0)$ and a true vacuum $(\phi=\phi_1)$. In this paper, we investigate the geodesic motion of both massive test particles and photons in the vicinity of HBH with  $\phi_1=0.5$ and $\phi_1=1.0$ by analyzing their effective potentials derived from the geodesic equation. By fixing $\phi_1$, the effective potential of a massive test particle increases monotonically when its angular momentum $L$ is very small. When $L$ increases to a critical value, the effective potential possesses an inflection point which is known as the innermost stable of circular orbit (ISCO), where the test particle can still remain stable in a circular orbit with a minimal radius without being absorbed by the HBH or fleeing to infinity. Beyond the critical value of $L$, the effective potential possesses a local minimum and a local maximum, indicating the existence of unstable and stable circular orbits, respectively. Moreover, the HBH possesses an unstable photon sphere but its location slightly deviates from the Schwarzschild black hole. The trajectories of null geodesics in the vicinity of HBH can also be classified into three types, which are the direct, lensing and photon sphere, based on the deflection angle of light, but the values of impact parameters can vary significantly than the Schwarzschild black hole.
\end{abstract}

\maketitle

\section{Introduction}

In General Relativity (GR), the no-hair theorem describes that the characteristics of a black hole can be completely determined by only three global charges: mass, electric charge, and angular momentum \cite{Israel:1967wq,Carter:1971zc,Ruffini:1971bza}, thus only electrovacuum black holes such as the Schwarzschild and Reissner-Nordstrom black holes can satisfy this theorem. However, there exists a class of black holes known as hairy black holes (HBHs), which manage to circumvent the no-hair theorem. The event horizon of these black holes are supported by nontrivial matter fields, allowing them to possess additional parameters referred to as "hair", which are associated to the corresponding theory. As a result, the solutions of HBHs can be bifurcated from the electrovacuum black holes, demonstrating unique features in the strong gravity regime, while becoming indistinguishable from them in the weak gravity regime. 

In particular, various types of asymptotically flat HBHs have been constructed numerically in the Einstein-Klein-Gordon (EKG) theory \cite{Corichi:2005pa,Chew:2022enh,Gubser:2005ih,Chew:2024rin,Chew:2023olq,Chew:2024evh}, where the Einstein gravity is minimally coupled to scalar potentials $V(\phi)$ without the inclusion of other matter fields. Hence, the existence of HBHs is fully determined by the characteristics of $V(\phi)$ to bypass the no-hair theorem \cite{Herdeiro:2015waa}. In general, $V(\phi)$ cannot be strictly positive. It should become negative in some regions in order to violate the weak energy condition - one of the basic criteria of the no-hair theorem. For instance, a HBH is constructed from one $V(\phi)$ of asymmetric vacua \cite{Corichi:2005pa, Chew:2022enh}, which has the false and true vacuums to describe the first-order phase transition of our universe \cite{Coleman:1980aw} and whose negative region is bounded by two roots of $V(\phi)$. Besides, that corresponding HBH has been generalized to become a charged HBH which can be bifurcated from the Reissner-Nordstrom black hole \cite{Chew:2024rin}. Furthermore, a HBH is constructed by $V(\phi)=-\lambda \phi^4 + \mu \phi^2$ with $\lambda$ and $\mu$ are the positive  constants, where the negative regions of $V(\phi)$ lies in the region $|\phi| > \sqrt{\mu/\lambda} $ \cite{Gubser:2005ih, Chew:2023olq}. The HBH can be connected smoothly with its counterpart gravitating scalaron in the small horizon limit \cite{Chew:2024bec}. Recently, a HBH can be constructed by $V(\phi)=-\alpha^2 \phi^6$ where the scalar field is massless, thus it serves as a first step to explore the possibilities of existence of HBH with generalized $V(\phi)=-\alpha^2 \phi^n$ with $n$ is an integer \cite{Chew:2024evh}. On the other hand, other constructions of HBHs in the EKG theory can be found in Refs. \cite{Bechmann:1995sa,Dennhardt:1996cz,Bronnikov:2001ah,Martinez:2004nb,Nikonov:2008zz,Anabalon:2012ih,Stashko:2017rkg,Gao:2021ubl,Karakasis:2023ljt,Atmaja:2023ing,Li:2023tkw,Rao:2024fox,Wijayanto:2022bwx,Wijayanto:2023wru}. Moreover, other classes of scalar HBHs can be constructed by minimally coupling a real scalar field with a U(1)-gauged complex scalar field \cite{Kunz:2023qfg,Kunz:2024uux,Brihaye:2024mlm} and Proca field associated with some potentials \cite{Herdeiro:2024pmv}.

So far GR has successfully passed numerous observational tests with remarkable precision since the groundbreaking proposal by A. Einstein at 1915,  ranging from the perihelion shift of mercury at 1920, the discovery of the binary pulsars by R. Hulse and J. Taylor \cite{Hulse:1974rfg}, and then  the recent detection of gravitational waves from the merger of binary compact objects by LIGO-VIRGO-KAGRA (LVK) collaboration and imaging of shadow for supermassive black holes in the galaxies M87 and Sgr A$^{*}$ by Event Horizon Telescope (EHT). Hence, these observational tests are actually the basic manifestation of GR about the notion of how the spacetime around a compact object can be bent by its own gravity and how does the corresponding spacetime affect the motion of a test particle, as summarized by the famous physicist J. Wheeler in his quotation ``Matter tells space how to curve. Space tells matter how to move". Therefore, we study the influence of nontrivial scalar field on the geodesic motion of test particles in the vicinity of HBH with asymmetric vacua \cite{Corichi:2005pa, Chew:2022enh} and report the results in this paper. Besides, the geodesic motion of test particles in the vicinity of other types of black holes have been extensively studied, for instance, Schwarzschild black hole \cite{Hagihara:1931abc,Witzany:2023bmq}, Schwarzschild-(A)dS black hole \cite{Hackmann:2008zza}, swirling universe \cite{Lim:2015oha,Lim:2020fnx,Capobianco:2023kse,Capobianco:2024jhe}, Myers-Perry black hole \cite{Diemer:2014lba}, regular black hole by Ay\'on-Beato and Grac\'ia \cite{Garcia:2013zud}, black hole in braneworld \cite{Liu:2024lda}, (2+1)-rotating black holes \cite{Soroushfar:2015dfz,Kazempour:2017gho}, SU(2)-(A)dS black hole in conformal gravity \cite{Hoseini:2016ztk}, supersymmetric $\hbox {AdS}_5$ black hole \cite{Drawer:2020mpw}, Kerr black hole \cite{Yang:2013yoa,Wei:2018aft,Liu:2023tcy,Cieslik:2023qdc}, Kerr-Newman black hole \cite{Yang:2013wya,Zhang:2017nhl,Wang:2022ouq,Li:2023bgn,Ko:2023igf}, $U(1)^2$ dyonic rotating black holes \cite{Flathmann:2016knq}, Kerr-Newman-Taub-NUT black hole \cite{Cebeci:2015fie}, black hole in the Einstein-Maxwell-dilaton-axion theory \cite{Flathmann:2015xia}, polymer black hole in the loop quantum gravity \cite{Chen:2024sbc}, black hole in the Einsteinian cubic gravity \cite{Li:2024tld}, Carrollian Reissner-Nordstrom black hole \cite{Chen:2024how}, Euclidean Schwarzschild geometry \cite{Battista:2022krl}. The trajectories of test particles in these black holes can be described analytically by the elliptic functions, since the closed form of these black holes are known. However, we can only numerically obtain the trajectories of test particles in this paper since we couldn't obtain the closed form of the HBH, which can only be constructed numerically by solving the Einstein equation.

This paper is organized as follows. In Sec.~\ref{sec:th}, we briefly introduce some basic theoretical setups to construct the HBH and properties of the HBH. In Sec~\ref{sec:geo}, we introduce the Lagrangian, derive the effective potential for the test particles around the HBH and then introduce our approach to calculate the trajectories of particles. In Sec~\ref{sec:res}, we present and discuss our numerical findings. Finally, in Sec.~\ref{sec:con}, we summarize our work and present an outlook.

\section{Scalar Hairy Black Holes with Asymmetric Vacua} \label{sec:th}

\begin{figure}
\centering
\mbox{
 \includegraphics[trim=50mm 170mm 20mm 20mm,scale=0.58]{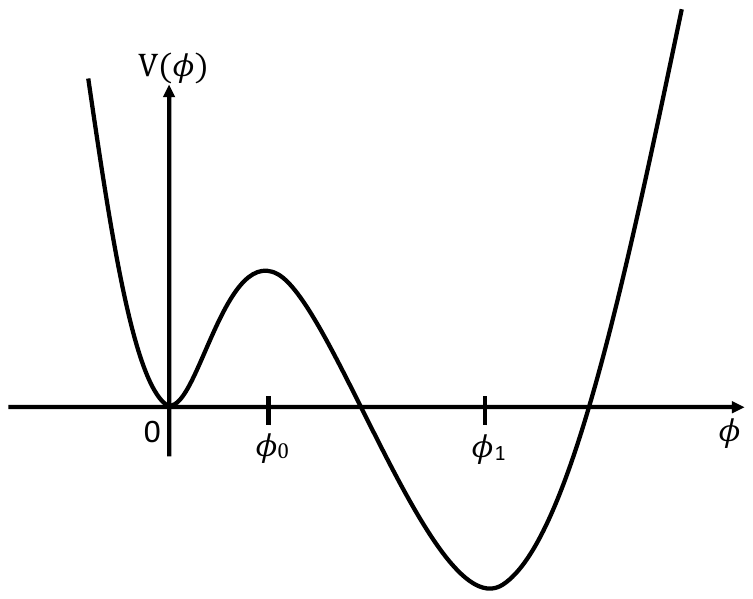}
%(b)
% \includegraphics[angle =0,scale=0.51]{Vsym}
}
\caption{The authors have considered the scalar potential $V(\phi)$ with a false vacuum at $\phi=0$, a barrier at $\phi=\phi_0$ and a true vacuum at $\phi=\phi_1$ to construct the hairy black holes \cite{Chew:2022enh}.}
\label{plot_Vphi}
\end{figure}

In the EKG theory, we consider a scalar potential $V(\phi)$ with asymmetric vacua \cite{Corichi:2005pa,Chew:2022enh} minimally coupled with the Einstein gravity,
\begin{equation} \label{EHaction}
 S=  \int d^4 x \sqrt{-g}  \left[  \frac{R}{16 \pi G} -  \frac{1}{2} \nabla_\mu \phi \nabla^\mu \phi - V(\phi) \right]  \,,
\end{equation}
where the explicit form of $V(\phi)$ is given by
\begin{equation} \label{Vpot}
 V(\phi) = \frac{V_0}{12} \left( \phi - a \right)^2 \left[ 3 \left( \phi-a\right)^2 - 4 (\phi-a) (\phi_0 + \phi_1) + 6 \phi_0 \phi_1  \right] \,,
\end{equation}
with $a$, $V_0$, $\phi_0$, and $\phi_1$ being constants. The asymptotic value of $\phi$ at the spatial infinity is fixed by $\phi=a$, and we choose $a=0$ such that $\phi=0$ at the spatial infinity in this paper. Hence, as shown in Fig.~\ref{plot_Vphi}, $V(\phi)$ possesses a local minimum with $V(0)=0$ when $\phi=0$. $V(\phi)$ also possesses a local maximum at $\phi=\phi_0$ and a global minimum at $\phi=\phi_1$. In the past decades $V(\phi)$ has been employed to study the first-order phase transition of our universe from the false vacuum $(\phi=0)$ to the true vacuum $(\phi=\phi_1)$. 

The negative region of $V(\phi)$  shown in Fig.~\ref{plot_Vphi} can violate the weak energy condition, which is one of the basic assumptions in the no-hair theorem. Hence, a few decades ago a class of HBHs can be constructed by Eq.~\eqref{Vpot} \cite{Corichi:2005pa} to evade the no-hair theorem, and recently its properties such as the Hawking temperature, Ricci scalar, Kretschmann scalar and so on just have been studied systematically \cite{Chew:2022enh}. Moreover, it can be bifurcated from the Reissner-Nordstrom black hole to possess an electric charge by minimally coupling the Maxwell field to the EKG theory \cite{Chew:2024rin}. Besides, Eq.~\eqref{Vpot} has been employed to construct a fermionic star and its tidal deformability just has been investigated recently \cite{DelGrosso:2023dmv,Berti:2024moe}.

Note that the asymmetrical profile of $V(\phi)$ is caused by the appearance of cubic term $\phi^3$, thus $V(\phi)$ can become symmetric if the cubic term disappears. For instance, the authors recently have employed a symmetric profile of $V(\phi)=-\lambda \phi^4 + \mu \phi^2$ ($\lambda$ and $\mu$ are the positive constants), which possesses two degenerate global maxima and a local minimum at $\phi=0$ to construct the HBH \cite{Chew:2023olq} and gravitating scalaron \cite{Chew:2024bec}. 
 
We then obtain the Einstein equation and Klein-Gordon (KG) equation from the variation of Eq.~\eqref{EHaction} with respect to the metric and scalar field, respectively
\begin{equation} 
 R_{\mu \nu} - \frac{1}{2} g_{\mu \nu} R = 8 \pi G \left(  -   \frac{1}{2} g_{\mu \nu} \nabla_\alpha \phi \nabla^\alpha \phi - g_{\mu \nu} V + \nabla_\mu \phi \nabla_\nu \phi    \right) \,,  \quad 
 \nabla_\mu \nabla^\mu \phi  =  \frac{d V}{d \phi} \,.  \label{eom}
\end{equation}

The following spherically symmetric metric is employed as an Ansatz to construct the numerical solutions of HBH,
\begin{equation}  \label{line_element}
ds^2 = - N(r) e^{-2 \sigma(r)} dt^2 + \frac{dr^2}{N(r)} + r^2  \left( d \theta^2+\sin^2 \theta d\varphi^2 \right) \,, 
\end{equation}
where $N(r)=1-2 m(r)/r$ with $m(r)$ is the Misner-Sharp mass function \cite{Misner:1964je}. The ADM mass of a HBH can be read off simply by the condition $m(\infty)=M$. The direct substitution of Eq. \eqref{line_element} into the Eq. \eqref{eom} yields a set of nonlinear ordinary differential equations (ODEs) for the functions,
\begin{equation}
m' = 2 \pi G \left( N \phi'^2 + 2 V \right) \,, \quad \sigma' = - 4 \pi G r \phi'^2 \,,    \quad
\left(  e^{- \sigma} r^2 N \phi' \right)' = e^{- \sigma} r^2  \frac{d V}{d \phi} \,,
\end{equation}
where the prime denotes the derivative of the functions with respect to the radial coordinate $r$. Here we only consider the solutions of HBH for the range of $r$ from the horizon radius $r_H$ to infinity. Although the form of ODEs looks very simple, it is almost impossible to obtain the analytical solutions for the HBH, hence we integrate the ODEs numerically with the appropriate boundary conditions by the professional ODE solver package Colsys \cite{Ascher:1979iha}, which adopts the Newton-Raphson method to solve the boundary value problem for a set of nonlinear ODEs by utilizing the adaptive mesh refinement to greatly enhance the accuracy of solutions with more than 1000 points, and estimating the errors of solutions for the users' reference. 

Here we impose the boundary conditions for the ODEs at the horizon and infinity. All functions are required to be finite at the horizon, which can be expressed in term of the power series expansion with few leading terms,
\begin{align}
 m(r) &= \frac{r_H}{2}+ m_1 (r-r_H) + O\left( (r-r_H)^2 \right) \,, \label{m_ex} \\
\sigma(r) &= \sigma_H + \sigma_1   (r-r_H) + O\left( (r-r_H)^2 \right)  \,, \\
 \phi(r) &= \phi_H +  \phi_{H,1}  (r-r_H) + O\left( (r-r_H)^2 \right)  \,,
\end{align} 
where
\begin{equation}
   m_1 = 4 \pi G r^2_H  V(\phi_H)  \,, \quad  \sigma_1 = -  4 \pi G r_H \phi^2_{H,1} \,, \quad   \phi_{H,1}= \frac{r_H \frac{d V(\phi_H)}{d \phi}}{1-8 \pi G r_H^2 V(\phi_H)}  \,, 
\end{equation} 
Here $\sigma_H$ and $\phi_H$ are the values of $\sigma(r), \phi(r)$ at the horizon respectively. The denominator of $\phi_{H,1}$ has to fulfill the condition $1-8 \pi G r_H^2 V(\phi_H) \neq 0$ such that $\sigma(r)$ and $\phi(r)$ are finite at the horizon. At  infinity all functions have to satisfy the asymptotically flat condition, which are given by $m(r \rightarrow \infty)=M$ with $M$ is the ADM mass of the HBH, $\sigma(r \rightarrow \infty)=\phi(r \rightarrow \infty)=0$. Furthermore, we perform the numerical calculation in the compactified coordinate $x \in [0,1]$ via the transformation $x=1-r_H/r$, which maps the horizon and infinity to the numerical value $0$ and $1$ respectively. Besides, we introduce the dimensionless parameters: $r \rightarrow  r/\sqrt{\beta}$, $m \rightarrow  m/\sqrt{\beta}$, $\phi \rightarrow \sqrt{\beta} \phi$, $\phi_1 \rightarrow \sqrt{\beta} \phi_1$, $\phi_0 \rightarrow \sqrt{\beta} \phi_0$ and $V \rightarrow \sqrt{\beta} V$. Therefore, the remaining free parameters in the calculation are $\phi_0$, $\phi_1$, $r_H$, $\sigma_H$, $\phi_H$, and $M$. The parameters $\sigma_H$ and $M$ can be determined exactly when the solutions of HBH satisfy the boundary conditions, so the input parameters for the calculation are given by $\phi_0$, $\phi_1$, $r_H$, and $\phi_H$.

\begin{figure}
\centering
\mbox{
(a)
 \includegraphics[angle =-90,scale=0.335]{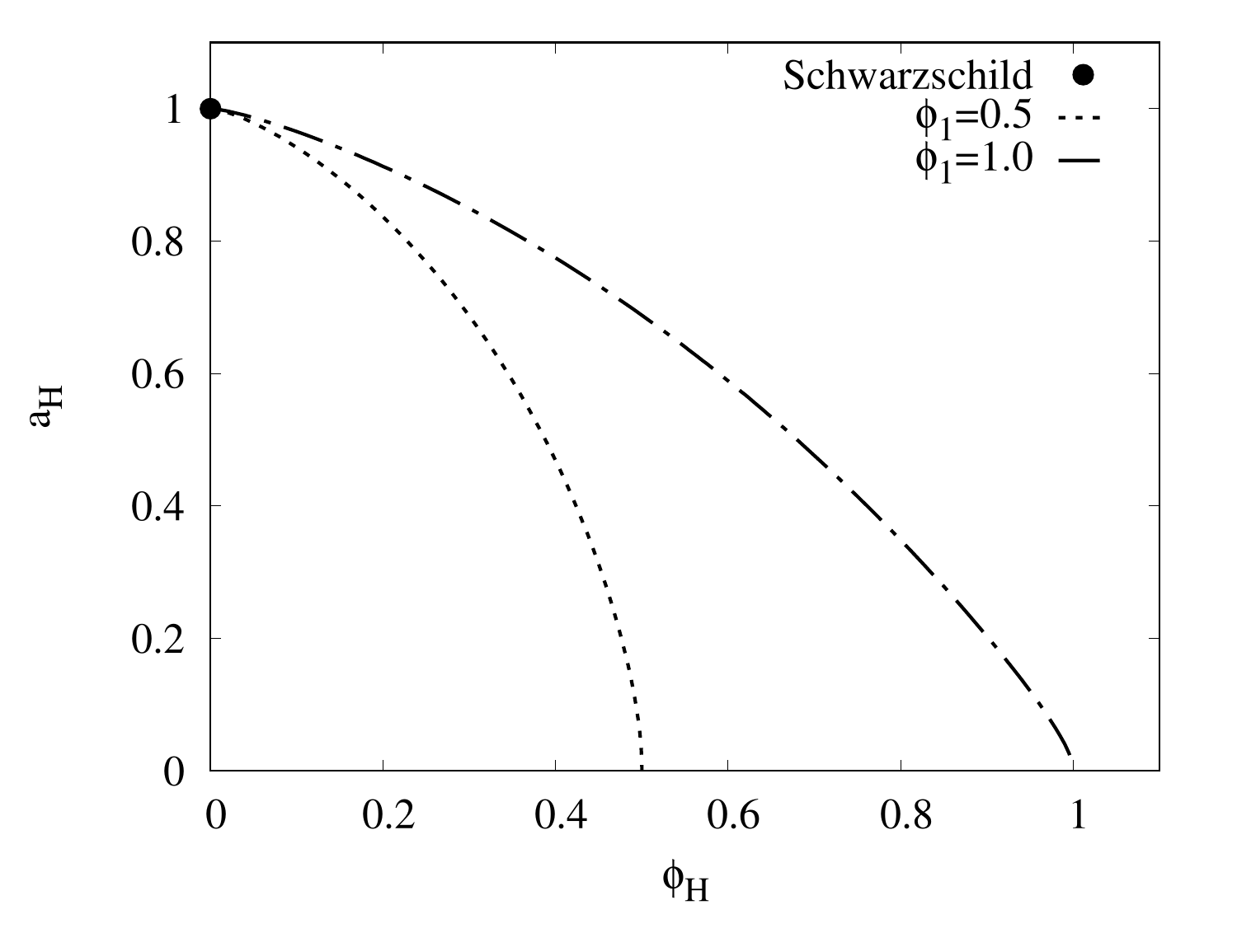}
(b)
\includegraphics[angle =-90,scale=0.335]{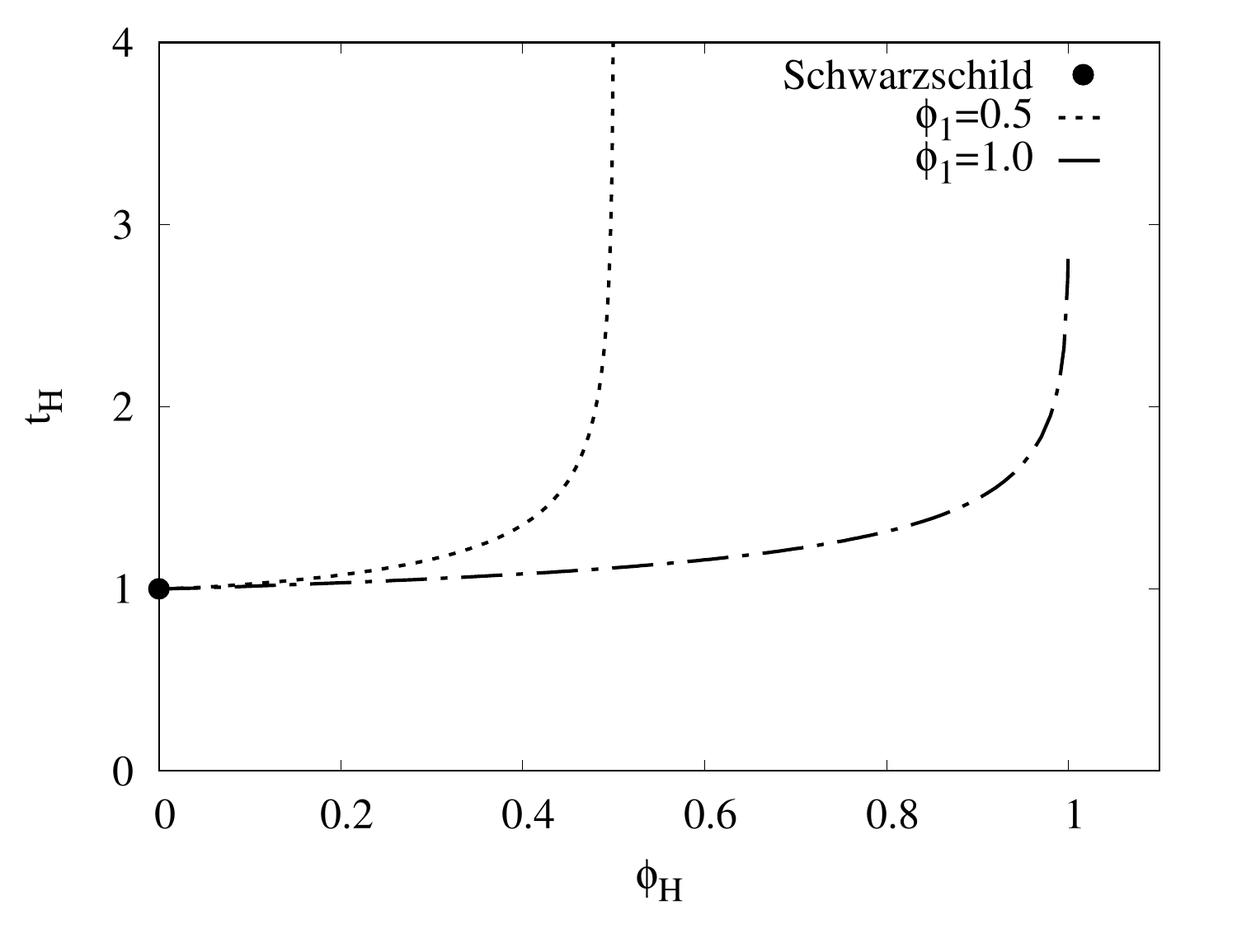}
}
\mbox{
(c)
 \includegraphics[angle =-90,scale=0.335]{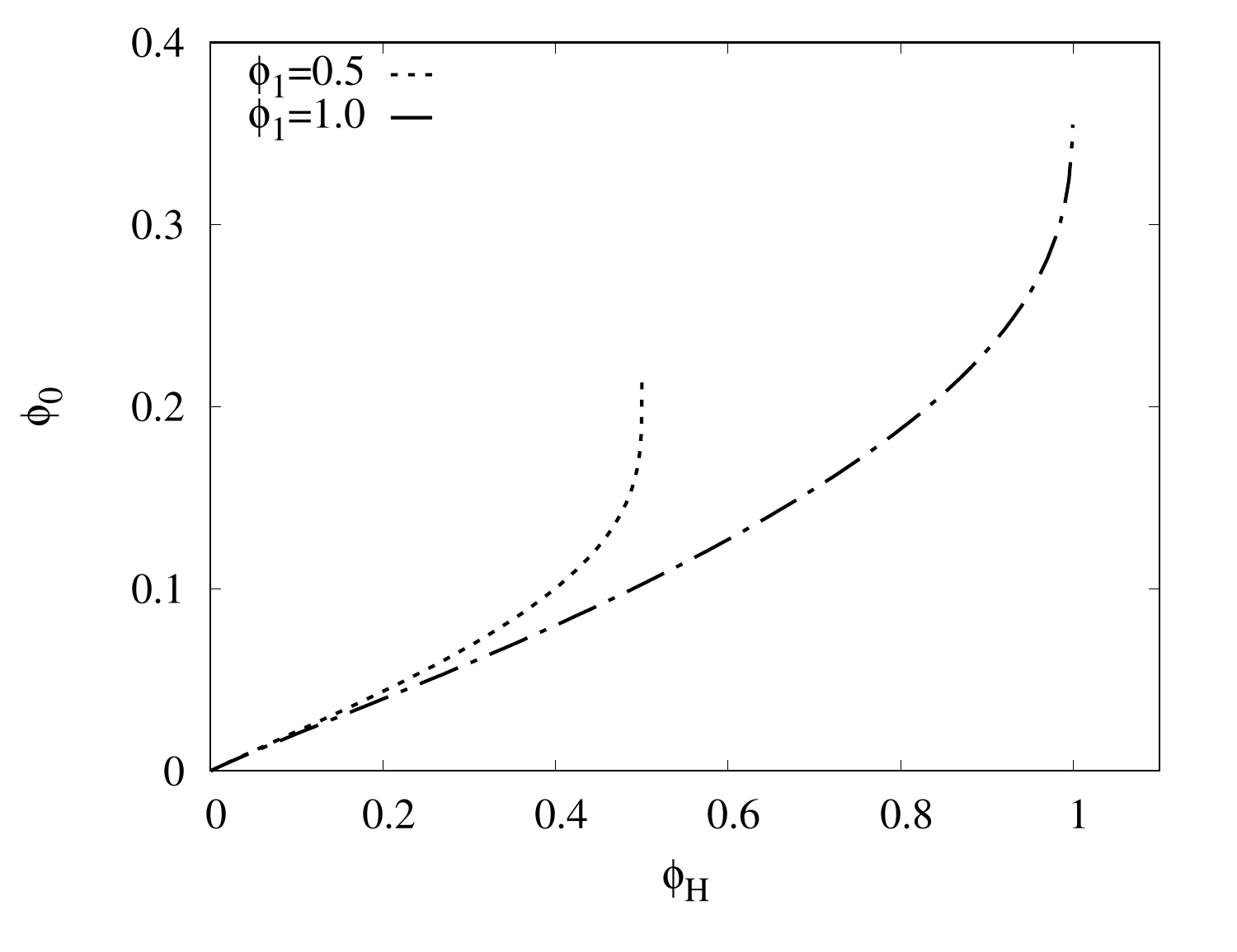}
 (d)
 \includegraphics[angle =-90,scale=0.335]{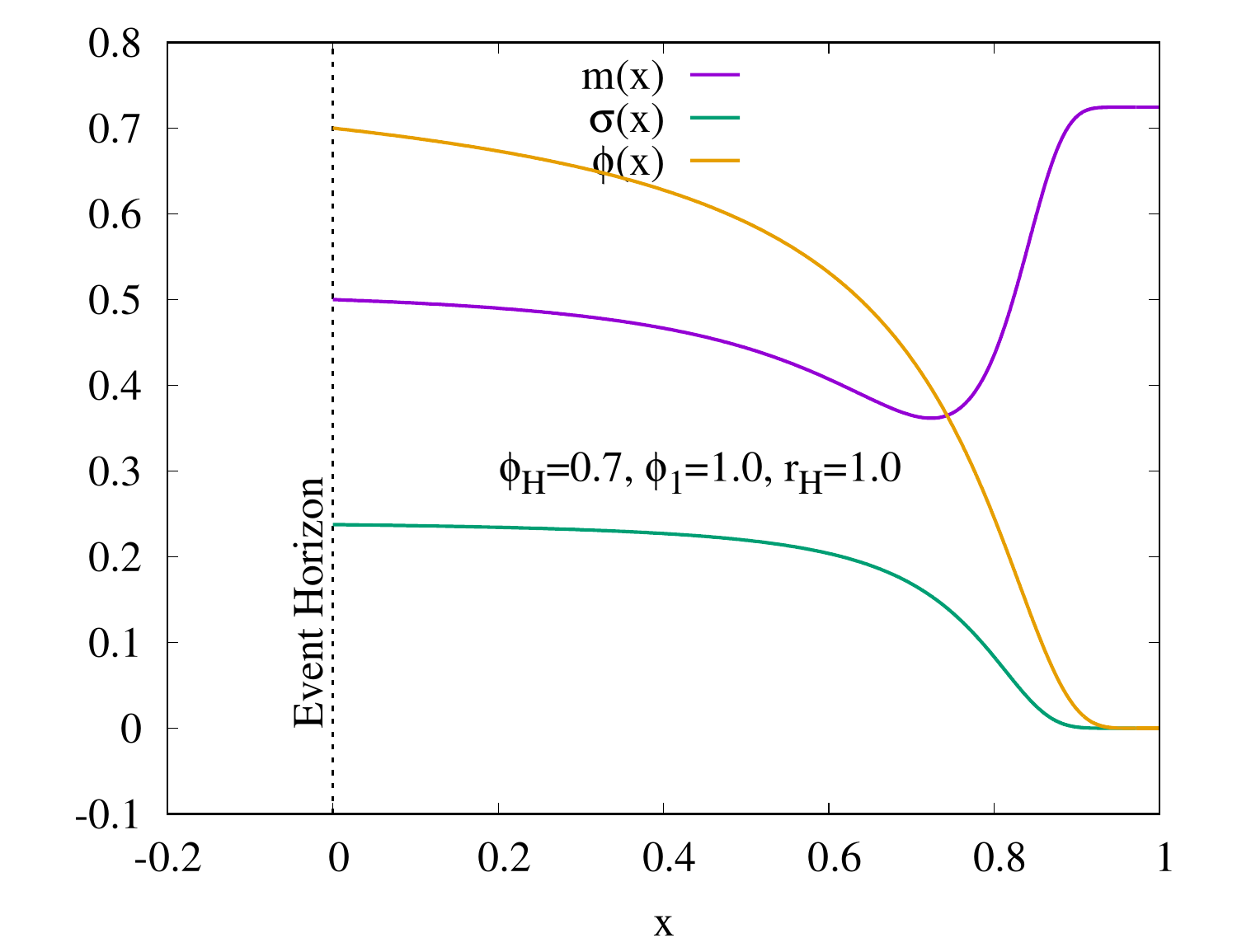}
}
\caption{Two basic properties of HBH with $\phi_1=0.5$ and $\phi_1=1.0$ when the scalar field $\phi_H$ is nontrivial at the horizon $r_H$ : (a) The reduced area of horizon $a_H$ vs $\phi_H$ (b) The reduced Hawking temperature vs $\phi_H$; (c) The parameter $\phi_0$ as the function of $\phi_H$ for $\phi_1=0.5$ and $\phi_1=1.0$; (d) The solutions $m(x)$, $\sigma(x)$ and $\phi(x)$ of the HBH with $\phi_H=0.7, \phi_H=1.0, r_H=1.0$ in the compactified coordinate $x$.}
\label{plot_HBH}
\end{figure}

Since we have 4 free parameters to describe our HBHs,  it would be convenient when we study the geodesics motion of test particles around the HBHs with $r_H=1$ by fixing $\phi_1=0.5, 1.0$ for simplicity. Hence, the basic properties of HBH can be briefly summarized in Fig.~\ref{plot_HBH}, where we have defined the reduced area of horizon $a_H=A_H/(16 \pi M^2)$ and reduced Hawking temperature $t_H=8 \pi T_H M$ with the area of horizon $A_H=4 \pi r^2_H$ and Hawking temperature $T_H = \frac{1}{4 \pi} N'(r_H) e^{-\sigma_H}$. The purpose to introduce such quantities is to easily see the connection between our HBH model with the Schwarzschild black hole, since the trivial solution to the Eq.~\eqref{eom} is the Schwarzschild black hole when the scalar field at the horizon $\phi_H$ doesn't exist. Recall that the values of $a_H$ and $t_H$ are unity for Schwarzchild black hole when $\phi_H=0$. However, when $\phi_H$ increases, the HBH bifurcates from the Schwarzschild black hole where $a_H$ decreases from unity but $t_H$ increases from unity. In the limit $\phi_H=\phi_1$, $a_H$ reaches to zero but $t_H$ increases very sharply, therefore the HBH could disappear in that limit.  

Figs.~\ref{plot_HBH}(c) demonstrates that $\phi_0$  increases monotonically as $\phi_H$ increases from zero for $\phi_1 = 0.5, 1.0$, and is almost indistinguishable for small $\phi_H$. We also find the values of $\phi_0  <<\phi_1$ in the limit $\phi_H=\phi_1$. Fig.~\ref{plot_HBH}(d) shows the typical profiles of the solutions of HBH with $\phi_H=0.7$, $\phi_1=1.0$ and $r_H=1$ in the compactified coordinate $x$. The functions behave almost constant inside the bulk, corresponding to the global minimum of $V(\phi)$ which is the true vacuum $\phi_1$. However, a sharp boundary developed at some intermediate region of the spacetime, when the solutions move away from the horizon, to drastically change to a new set of almost constant functions. This region corresponds to the false vacuum $(\phi=0)$ at infinity, where $\phi$ sits at the local minimum of $V(\phi)$.

\section{Geodesic Motion of Test Particles Around the HBH} \label{sec:geo}

The geodesic motion of test particles in the vicinity of HBH can be studied in the Lagrangian formalism, where the Lagrangian is given by
\begin{equation}
 \mathcal{L} = \frac{1}{2} \dot{x}^\mu \dot{x}_\mu = \epsilon \,,
\end{equation} 
where $\epsilon=0$ refers to massless particle and $\epsilon=-1$ refers to massive particle. The dot denotes the derivative of a function with respect to the affine parameter $\lambda$. Since the spacetime of HBH is static and stationary, it possesses two conserved quantities which are the energy $E$ and angular momentum $L$, 
\begin{equation}
 E = - \frac{\partial \mathcal{L} }{\partial \dot{t}} = e^{-2 \sigma} N \dot{t} \,, \quad  L =  \frac{\partial \mathcal{L}}{\partial \dot{\varphi}} = r^2 \dot{\varphi} \,. 
\end{equation}
Since our HBH is spherically symmetric, then we just have to concentrate on the motion of particles on the equatorial plane $(\theta=\pi/2)$. Therefore, we can obtain the radial equation $\dot{r}$ as shown in the below,
\begin{equation}
 e^{-2 \sigma} \dot{r}^2 =    E^2 -  V_{\text{eff}}(r)  \,, 
\end{equation}
where the effective potential $V_{\text{eff}}$ of a test particle is given by
\begin{equation}
 V_{\text{eff}}(r) = e^{-2 \sigma} N \left(  \frac{L^2}{r^2} - \epsilon \right) \,.
\end{equation}

There are several types of orbits can exist in the spacetime of HBH by analyzing $V_{\text{eff}}(r)$, such as the circular orbit, bound orbit, escape orbit and etc. In general, a test particle is in a circular orbit when it is moving in a circular motion with a fixed radius from the HBH. It can be determined from the condition $V'_{\text{eff}}(r)=0$. If $V_{\text{eff}}(r)$ possesses a local minimum, then the circular orbit is stable but the circular orbit is unstable if $V_{\text{eff}}(r)$ possesses a local maximum. A test particle is in a bound orbit where it can travel between the minimal radius and maximal radius from the HBH with $E<1$. However, a test particle can flee to the infinity from a point in HBH when $E>1$. If $V_{\text{eff}}(r)$ contains an inflection point which connects the curves with concave downward and concave upward, this implies the test particle is located in the innermost stable circular orbit (ISCO), which can be determined by the condition $V'_{\text{eff}}(r)=V''_{\text{eff}}(r)=0$, and the explicit form is given by
\begin{align}
  V'_{\text{eff}}(r) &=  \left( \frac{N'}{N} - 2 \sigma ' \right)  V_{\text{eff}} - \frac{2 L^2 e^{-2 \sigma } N}{r^3}    \,, \\
 V''_{\text{eff}}(r) &= \left(  \frac{N''}{N} - \frac{4 \sigma' N'}{N} + 4 N \sigma'^2 - 2 \sigma''  \right)    V_{\text{eff}} + 2 L^2 e^{-2\sigma} \left( \frac{3 N}{r^4} + \frac{4 N \sigma'}{r^3} - \frac{2 N'}{r^3}  \right) \,.
\end{align}
The location of ISCO, $r_{\text{ISCO}}$ can be calculated numerically by solving the above equations. 

Moreover, we derive the following expression,
\begin{equation}
 \frac{d \varphi}{dr} = \frac{ \dot{\varphi} }{ \dot{r} } = \frac{L}{r^2 e^{ \sigma} \sqrt{  E^2 -  V_{\text{eff}}(r) }} \,,
\end{equation}
which can allow us to obtain the trajectories of test particles by numerically integrating the above expression, as shown in the below,
\begin{equation}
 \varphi(r) = \int_{r_{\text{min}}}^{r_{\text{max}}} \frac{L}{r^2 e^{ \sigma} \sqrt{  E^2 -  V_{\text{eff}}(r) }} dr  \,.\label{dvarphi}
\end{equation}
The above integration range $[r_{\text{min}},r_{\text{max}}]$ is only valid when $E^2 -  V_{\text{eff}}(r) \geq 0$. Then we can present their trajectories in the two-dimensional Cartesian coordinate using the following relation,
\begin{equation}
 x = r \cos( \varphi(r) ) \,, \quad y = r \sin( \varphi(r) ) \,.
\end{equation}

For photon $(\epsilon=0)$, we can introduce the parameter $b \equiv L/E$ which is known as the impact parameter in the radial equation $\dot{r}$,
\begin{equation}
 e^{-2 \sigma} \dot{r}^2 =  E^2 \left( 1 - b^2  v_{\text{eff}}(r) \right) \,,
\end{equation}
where 
\begin{equation}
 v_{\text{eff}}(r) =  \frac{e^{-2 \sigma} N}{r^2} \,.
\end{equation}
The extremum of $v_{\text{eff}}(r)$ indicates the presence of a photon sphere, where the null geodesics can theoretically orbit arbitrary amount of time. Since the imaging of shadow for the supermassive black holes in the center of some galaxies can also be described by the framework of backward ray-tracing, therefore it might be convenient to show the trajectories of null geodesics around the HBH according to the total number of orbits of null geodesics $n$, which can be defined as \cite{Gralla:2019xty}
\begin{equation}
 n =  \frac{\varphi}{2\pi} \,.
\end{equation} 
This allows us to classify the trajectories of null geodesics into three types: direct emission with $0<n<3/4$ for $b \in \left(0, b_2^{-} \right) \cup \left(b_2^{+}, \infty \right)$ where the null geodesics only intersect with the accretion disk once $(m=1)$; lensing with $3/4<n<5/4$ for $ b \in \left(b_2^{-}, b_3^{-} \right) \cup \left(b_3^{+}, b_2^{+} \right)$ where the null geodesics intersect with the accretion disk twice $(m=2)$; and photon sphere with $n>5/4$ for $b \in \left(b_3^{-}, b_3^{+} \right)$ where the null geodesics intersect with the accretion disk three times $(m=3)$. Note that $b_m^{-} < b_c$ and $b_m^{+} > b_c$, where $m$ represents the number of intersections of null geodesics with the accretion disk and $b_c$ is the critical value of $b$ for the photon sphere.

\section{Results and Discussions}\label{sec:res}

\begin{figure}
\centering
\mbox{
(a)
 \includegraphics[angle =-90,scale=0.335]{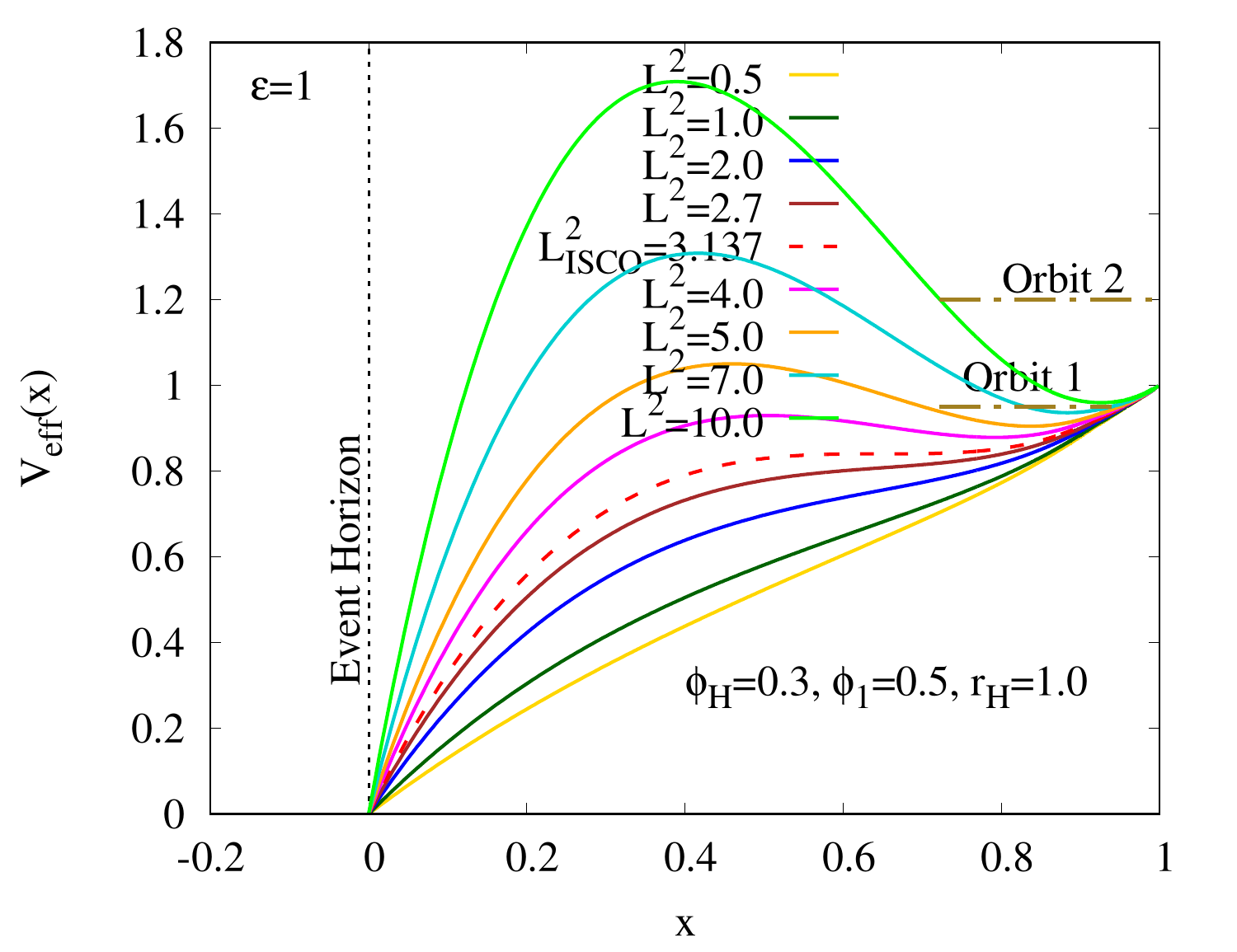}
(b)
 \includegraphics[angle =-90,scale=0.335]{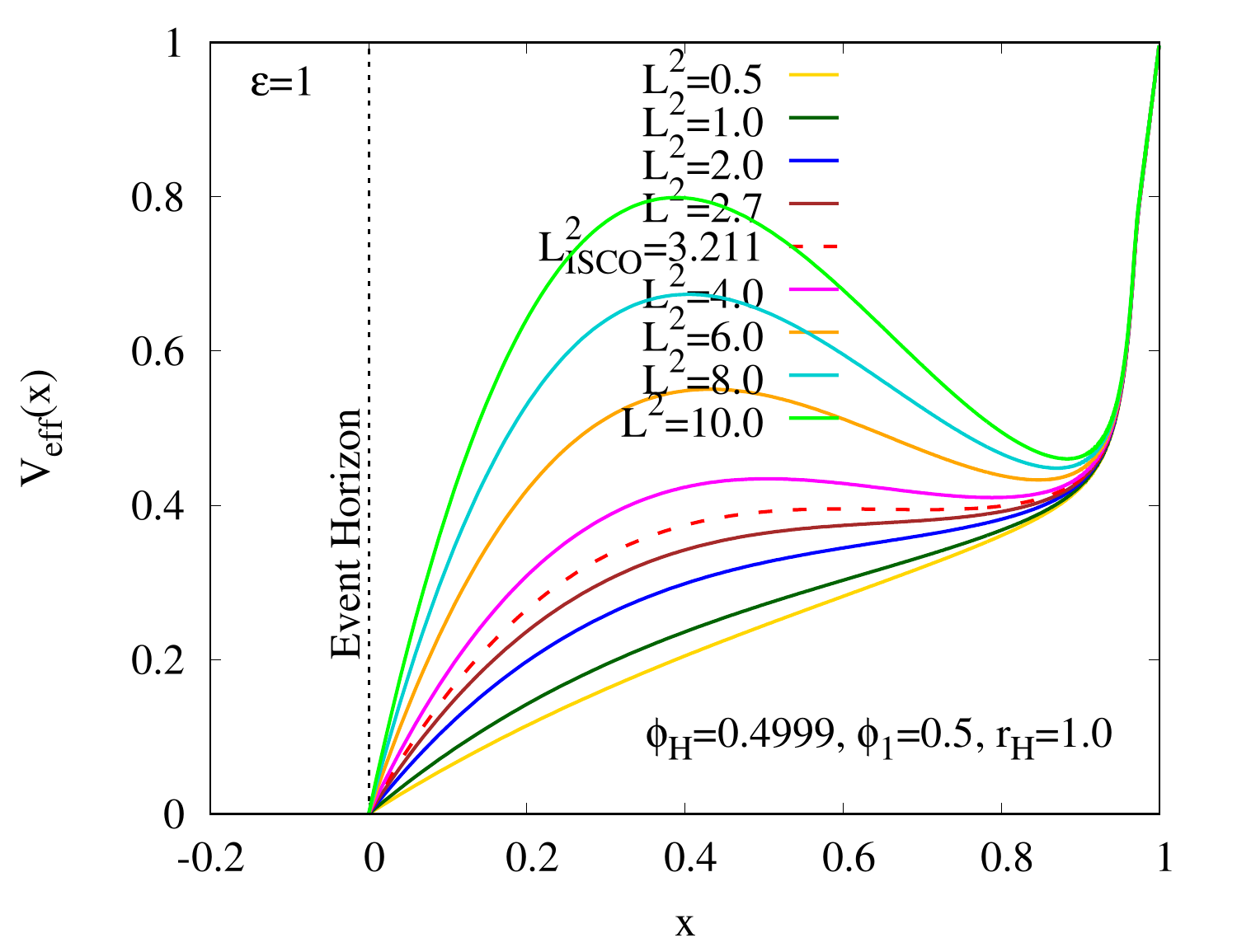}
 }
\mbox{
(c)
\includegraphics[angle =-90,scale=0.335]{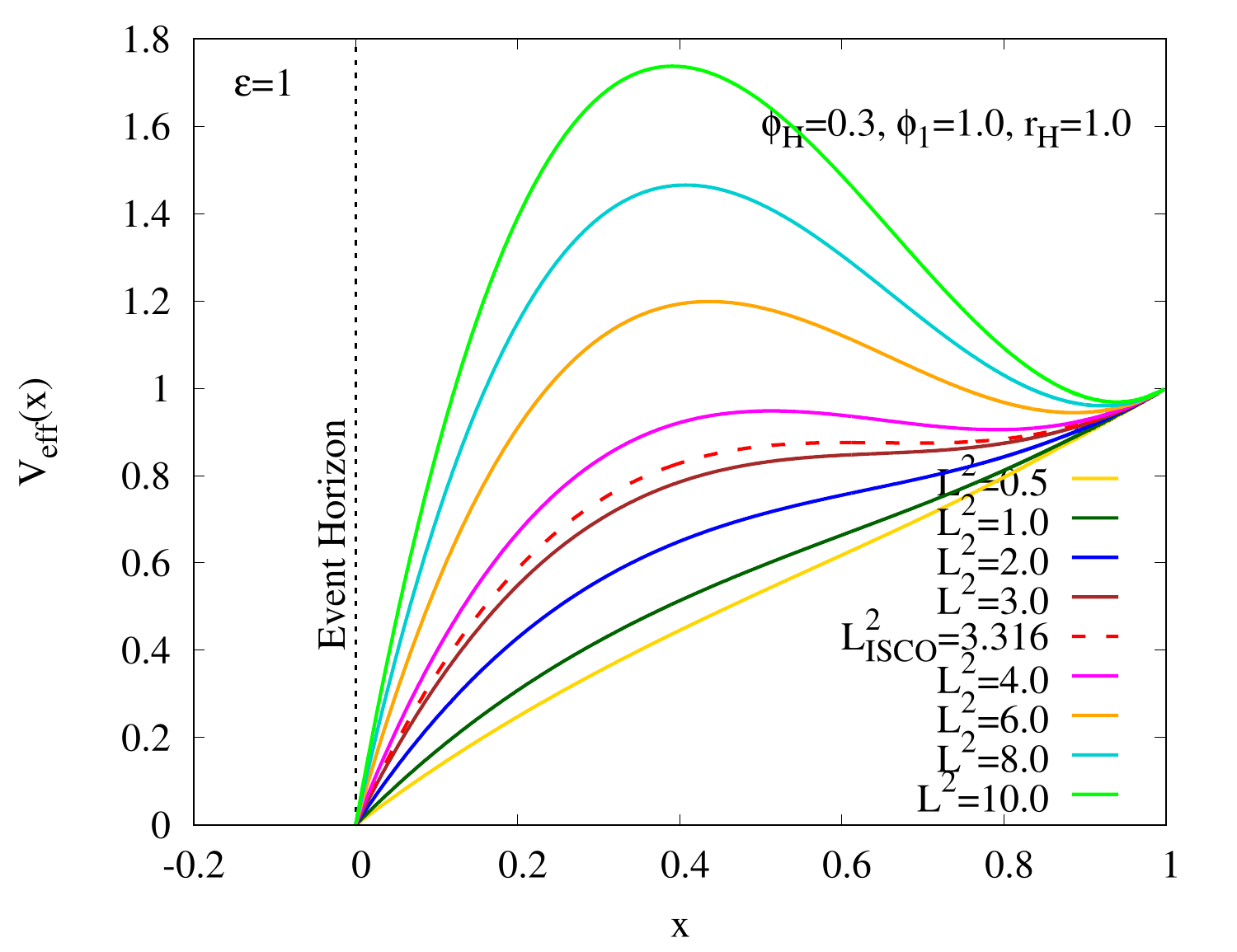}
(d)
 \includegraphics[angle =-90,scale=0.335]{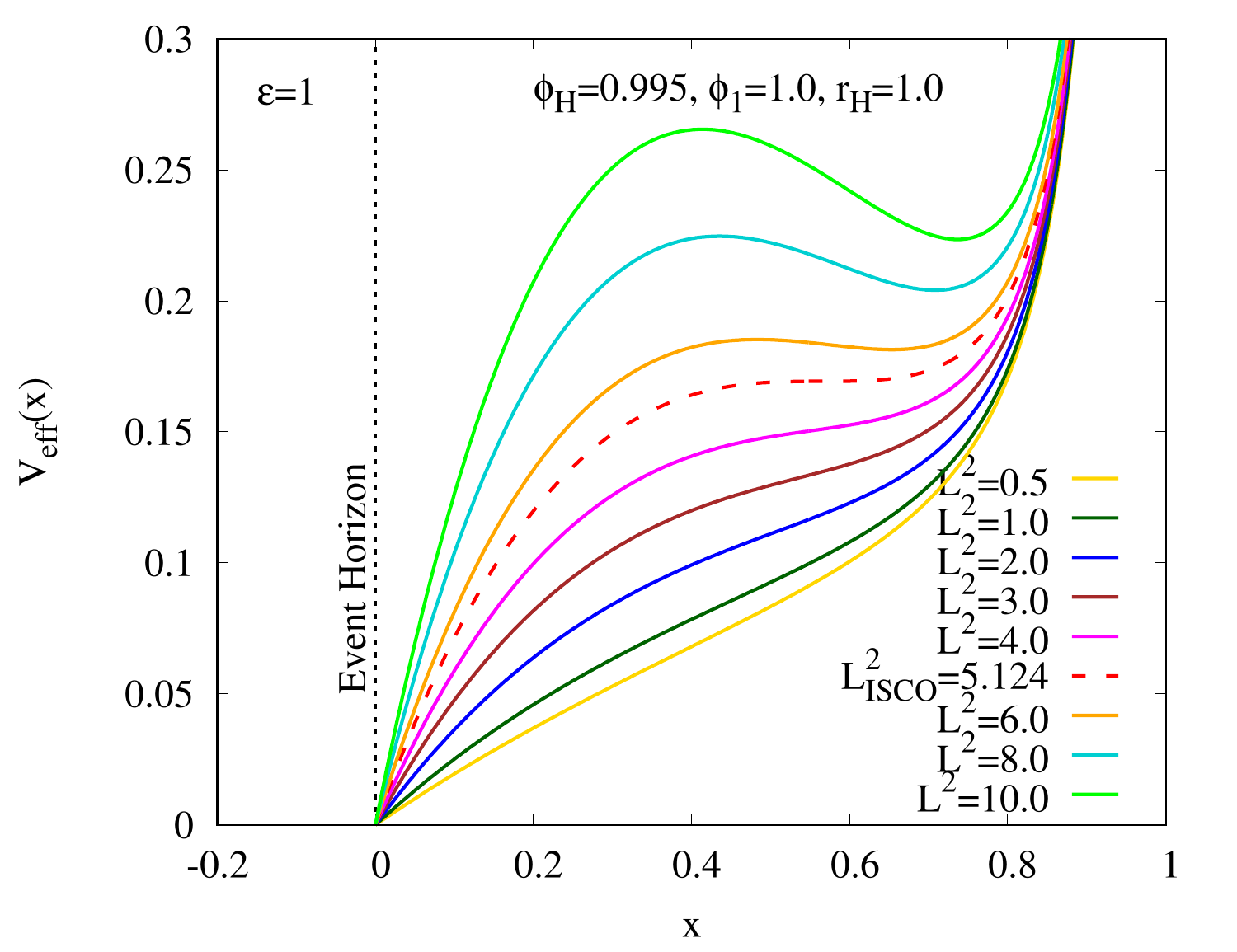}
}
\mbox{
(e)
\hskip21pt
 \includegraphics[angle =0,scale=0.78]{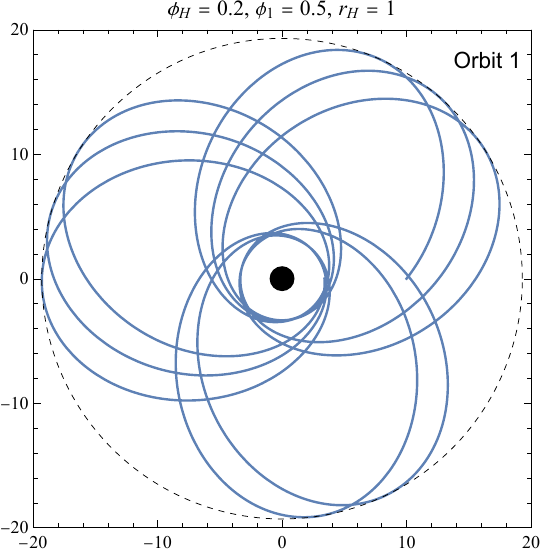} 
 \hskip10pt
(f)
\hskip30pt
 \includegraphics[angle =0,scale=0.78]{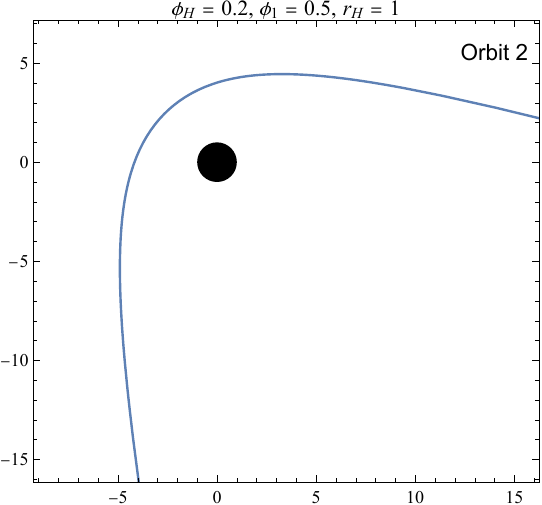}
 }
\caption{The effective potential $V_{\text{eff}}(x)$ in the compactified coordinate $x=1-r_H/r$ for a massive particle around the HBH with $r_H=1$ for a) $\phi_1=0.5$ and $\phi_H=0.3$. Note that in the figure, ``Orbit 1" and ``Orbit 2" refer to the motion of particle with $E^2=0.95$ in (e) and $E^2=1.2$ in (f), respectively; b) $\phi_1=0.5$ and $\phi_H=0.4999$; c) $\phi_1=1.0$ and $\phi_H=0.3$; d) $\phi_1=1.0$ and $\phi_H=0.995$; e) The bound orbit of a massive particle with $E^2=0.95$, labeled by ``Orbit 1" in (a) around the HBH with $\phi_1=0.5$, $\phi_H=0.3$, $r_H=1$; f) A massive particle with $E^2=1.2$, labeled by ``Orbit 2" in (a) fleeing away from the HBH with $\phi_1=0.5$, $\phi_H=0.3$, $r_H=1$. Note that the black disc represents the HBH.}
\label{plot_Vmassive1}
\end{figure}

We present our research findings for the investigation of geodesic motion of test particles around the HBH with $r_H=1$ by fixing $\phi_1=0.5$ and $\phi_1=1.0$. Thus, we begin with the massive test particle by analyzing its effective potentials $V_{\text{eff}}(x)$ with several values of the angular momentum squared $L^2$ in the compactified coordinate $x=1-r_H/r$ as shown in Figs.~\ref{plot_Vmassive1}(a) for $\phi_H=0.3$ and $\phi_1=0.5$. When we increase $L^2=0$ until $L^2_{\text{ISCO}}=3.137$ (red dashed curve), $V_{\text{eff}}(x)$ changes its monotonically increasing behaviour until possessing an inflection point, which is known as the location of ISCO, denoted as $x_{\text{ISCO}}$ when $L^2_{\text{ISCO}}=3.137$, causing $V'_{\text{eff}}(x_{\text{ISCO}})=V''_{\text{eff}}(x_{\text{ISCO}})=0$. This implies that the massive test particle can stay in the ISCO without being absorbed by HBH or escaped to infinity. When $L^2 > 3.137$, $V_{\text{eff}}(x)$ starts to develop a local maximum and a local minimum which indicate the massive test particle possesses unstable and stable circular orbits, respectively. Note that the location of unstable circular orbit is always less than the location of stable circular orbit, possibly due to the strong gravitational pull by the HBH causing the instability of the circular orbit in the bulk region. 

Fig.~\ref{plot_Vmassive1} (b) shows that $V_{\text{eff}}(x)$ of a massive test particle around the HBH with $\phi_H=0.4999$ (in the limit  $\phi_H = \phi_1$), which also behaves qualitatively similar to Fig.~\ref{plot_Vmassive1}(a). It possesses the ISCO when we increase $L^2$ from zero until $L^2_{\text{ISCO}}=3.221$ and two circular orbits which are stable and unstable when $L^2>3.221$. Recall that the values of $x_\text{ISCO}=2/3 \approx 0.6667 $ \cite{schval1} and $L^2_{\text{ISCO}}=3r^2_H$ for the Schwarzschild black hole, hence we find that $x_\text{ISCO}$ and $L^2_{\text{ISCO}}$ of the HBH with $\phi_1=0.5$ don't deviate too much from the Schwarzschild black hole. Other values of $x_\text{ISCO}$ and $L^2_{\text{ISCO}}$ correspond to different $\phi_H$ can be found in Table.~\ref{ta1}.

\begin{table}[H]
\begin{center}
\begin{tabular}{ |c|c|c|c|c|c| } 
 \hline
 $\phi_H$ & 0 (Schwarzschild)  & 0.2  &  0.4 & 0.45 & 0.4999    \\ 
\hline
 $x_{\text{ISCO}}$  &  0.6667  &  0.6607 
&  0.6428 & 0.6398 & 0.6509  
 \\
 \hline
 $L^2_{\text{ISCO}}$ & 3.0000    & 3.0545 
      & 3.2200  & 3.2340  & 3.2110 
\\   
 \hline
\end{tabular}
\caption{Several values of $x_{\text{ISCO}}$ and $L^2_{\text{ISCO}}$ correspond to a massive test particle around the HBH with $r_H=1$ and $\phi_1=0.5$. \label{ta1}}
\end{center} 
\end{table}

Figs.~\ref{plot_Vmassive1} (c) and (d) exhibit $V_{\text{eff}}(x)$ of a massive test particle around the HBH with $\phi_1=1.0$ for $\phi_H=0.3$ and $\phi_H=0.995$ (in the limit  $\phi_H = \phi_1$), respectively. We observe that the profiles of $V_{\text{eff}}(x)$ in these two figures also behave qualitatively similar to Figs.~\ref{plot_Vmassive1}(a) and (b), where they possess the ISCO when $L^2=L^2_{\text{ISCO}}$ and two circular orbits which are stable and unstable when $L^2>L^2_{\text{ISCO}}$. However, we find that the deviation of $L^2_{\text{ISCO}}$ and $x_{\text{ISCO}}$ of the HBH from the Schwarzschild black hole can become larger  when $\phi_H$ increases, as shown in Table.~\ref{ta2}. Besides, Figs.~\ref{plot_Vmassive1} (a)-(d) show that $V_{\text{eff}}$(x) approaches to unity when $x$ approaches to infinity, since the HBH is asymptotically flat.

\begin{table}
\begin{center}
\begin{tabular}{ |c|c|c|c|c|c|c|c| } 
 \hline
 $\phi_H$ & 0 (Schwarzschild)  & 0.2  &  0.4 & 0.6 & 0.9  & 0.95  & 0.995    \\ 
\hline
 $x_{\text{ISCO}}$  & 0.6667   & 0.6624
 & 0.6680 & 0.6828 & 0.5731 & 0.5644
& 0.5671\\
 \hline
 $L^2_{\text{ISCO}}$ & 3.0000  & 3.0988
& 3.5276 & 4.5204 & 5.7238 & 5.6721 & 5.1242\\   
 \hline
\end{tabular}
\caption{Several values of $x_{\text{ISCO}}$ and $L^2_{\text{ISCO}}$ correspond to a massive test particle around the HBH with $r_H=1$ and $\phi_1=1.0$. \label{ta2}}
\end{center} 
\end{table}

Moreover, Fig.~\ref{plot_Vmassive1} (e) demonstrates that a massive test particle with energy $E^2=0.95$, labeled by ``Orbit 1" in Fig.~\ref{plot_Vmassive1} (a) can move in a bound orbit around the HBH with $\phi_1=0.5$, $\phi_H=0.3$ and $r_H=1.0$. Nevertheless, it can flee to infinity with $E^2=1.2>1$, labeled by ``Orbit 2" in Fig.~\ref{plot_Vmassive1} (a) from the same HBH, as shown in Fig.~\ref{plot_Vmassive1} (f). These two figures can be obtained by numerically integrating Eq.~\eqref{dvarphi}.

\begin{figure}
\centering
\mbox{
(a)
 \includegraphics[angle =-90,scale=0.335]{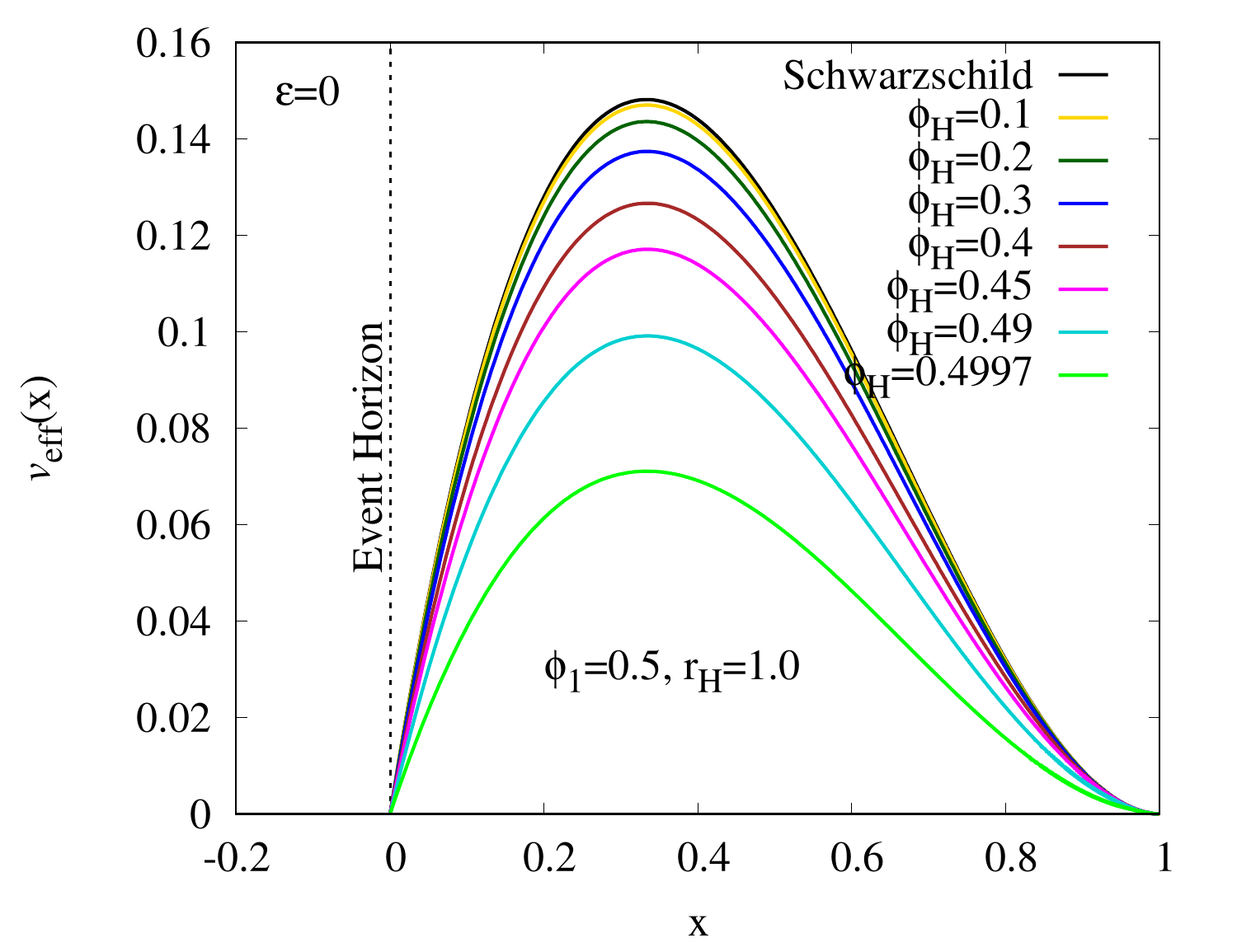}
(b)
\includegraphics[angle =-90,scale=0.335]{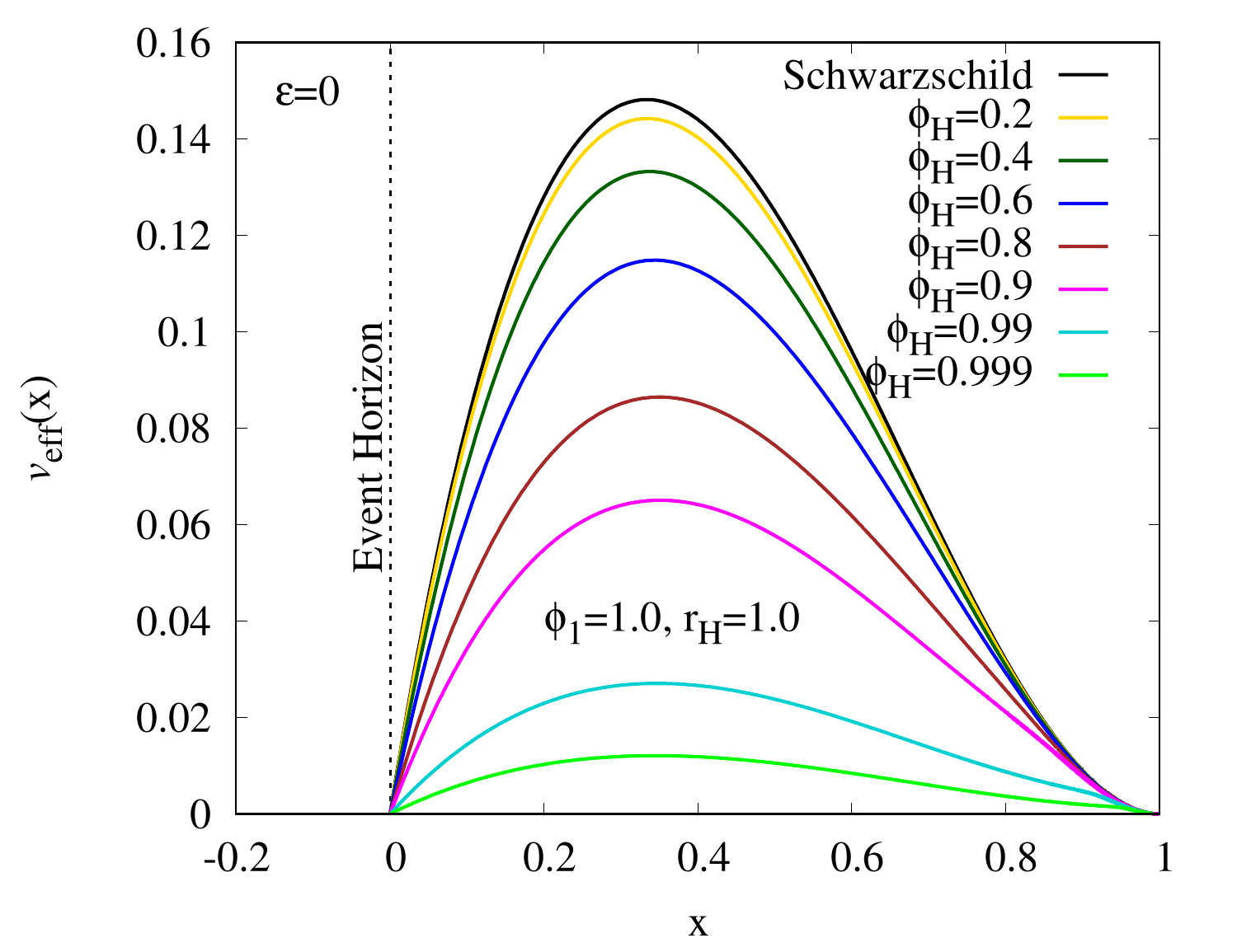}
}
\mbox{
(c)
 \includegraphics[angle =-90,scale=0.335]{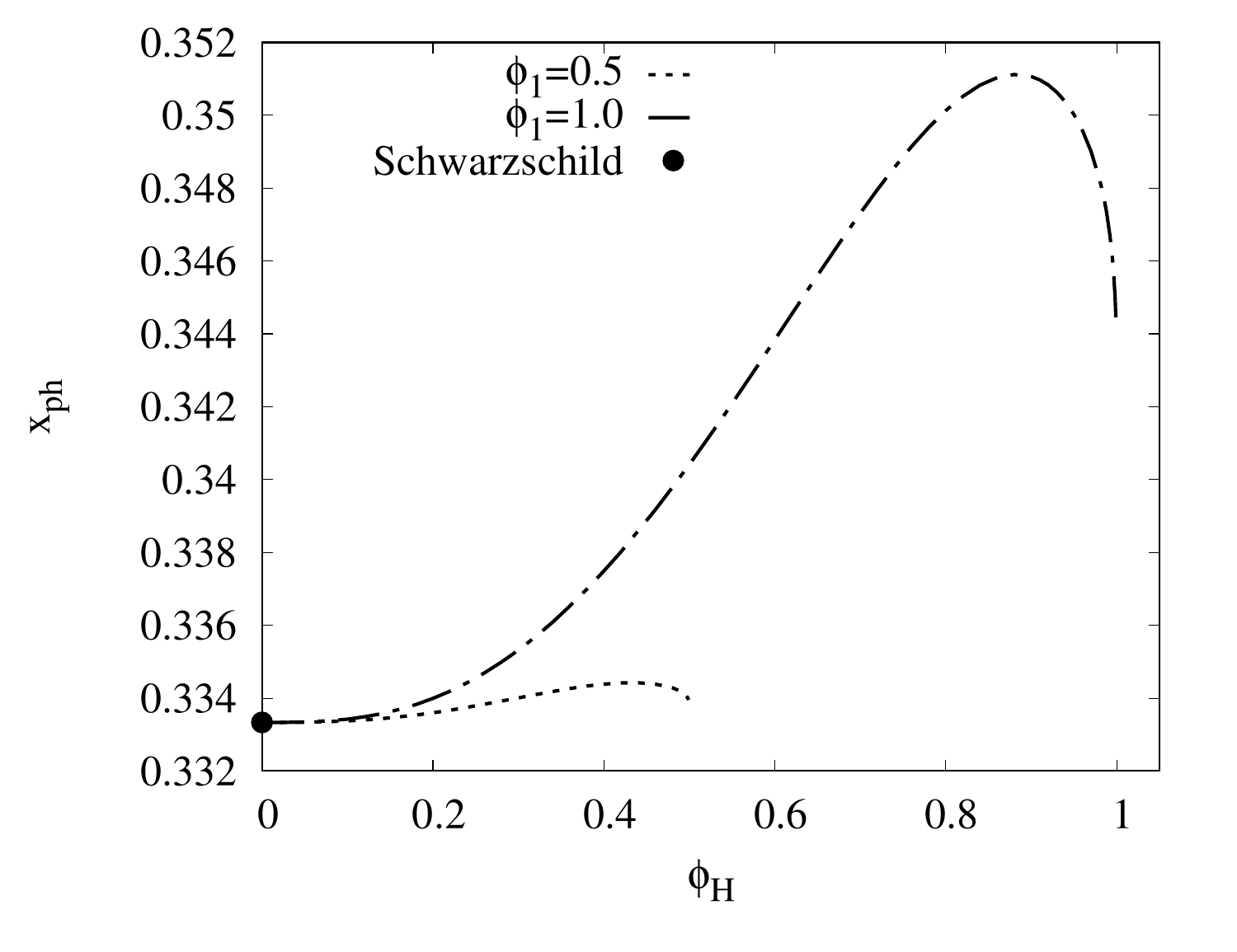}
(d)
\includegraphics[angle =-90,scale=0.335]{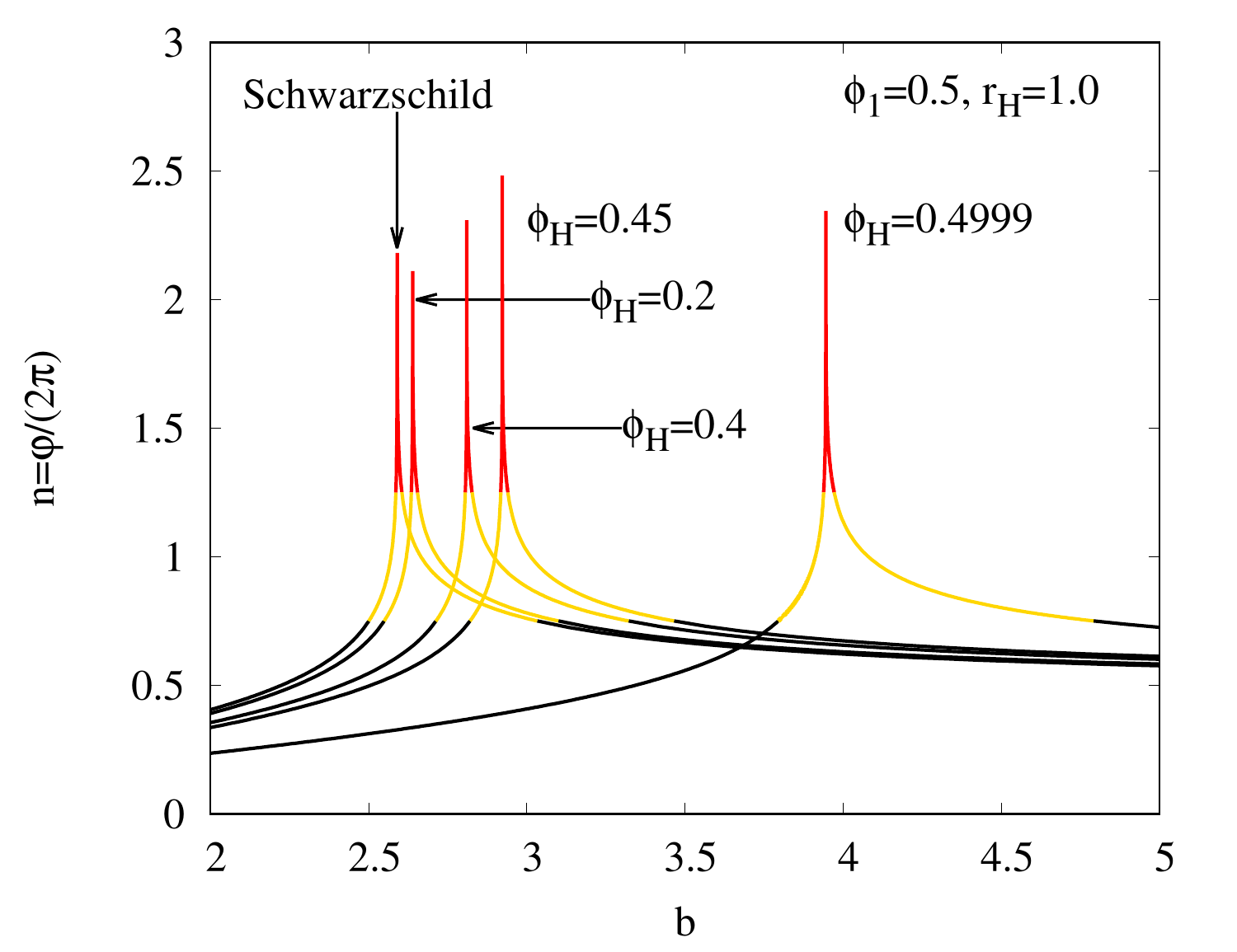} 
 }
\mbox{
\hskip-220pt
(e)
\includegraphics[trim = 0mm 0mm 0mm 0mm, clip,scale=0.72]{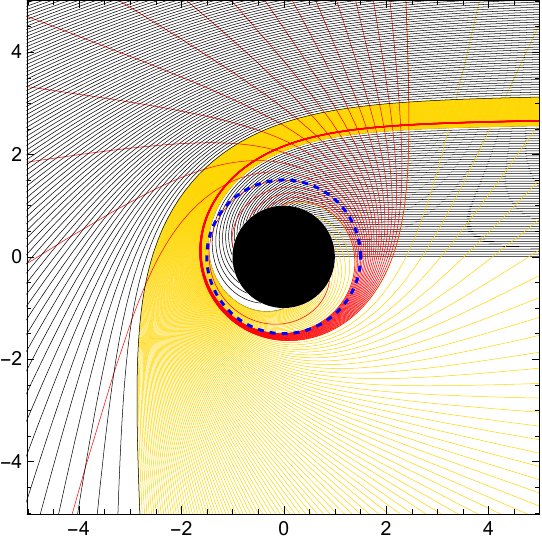} 
}
\caption{The effective potential $v_{\text{eff}}(x)$ in the compactified coordinate $x=1-r_H/r$ for null geodesics around the HBH with $r_H=1$ for a) $\phi_1=0.5$; b) $\phi_1=1.0$; c) The location of photon sphere in the compactified coordinate $x$ around the HBH with $r_H=1$, $\phi_1=0.5$ and $\phi_1=1.0$ versus $\phi_H$; 
d) Based on $n=\varphi/(2\pi)$, the null geodesics around the Schwarzschild black hole with $r_H=1$ and HBH with $\phi_1=0.5$, $r_H=1$ can be categorized into three emissions as a function of impact parameter $b$: direct (black), lensing (yellow) and photon ring (red).
e) The trajectories of null geodesics consist of direct (black curves), lensing (yellow curves) and photon sphere (blue curve) around the HBH with $\phi_1=0.5$, $\phi_H=0.3$, $r_H=1$.
}
\label{plot_Vph}
\end{figure}

Fig.~\ref{plot_Vph} (a) and (b) shows the effective potential $v_{\text{eff}}(x)$ of the null geodesics in the compactified coordinate $x$ in the HBH with $\phi_1=0.5$ and $\phi_1=1.0$, respectively. Both figures depict that $v_{\text{eff}}(x)$ possesses a local maximum, which indicates that the photon sphere is unstable. Recall that the location of photon sphere for the Schwarzchild black hole in the compactified coordinate $x$ is given by $x_{\text{ph}}=1/3$ \cite{schval2}, thus Fig.~\ref{plot_Vph}(c) shows the location of photon sphere for HBH with $\phi_1=0.5$ and $\phi_1=1.0$ don't deviate too much from the Schwarzschild black hole. Fig.~\ref{plot_Vph} (d) compares the total number of orbit $n$ for the Schwarzschild black hole and the HBH with $\phi_1=0.5$. Although the trajectories of null geodesics around the HBH still can be divided into three types: direct, lensing and photon sphere, we observe that when $\phi_H$ increases, the values of impact parameters $b_2^{-}$, $b_2^{+}$, $b_3^{-}$ and $b_3^+
$ associated to different types of trajectories as shown in Table \ref{ta3} for the HBH can increase significantly from the Schwarzcshild black hole, this can impact the brightness and size of the optical appearance (shadow) of the HBH. Besides, the three types of trajectories of null geodesics around the HBH with $\phi_1=0.5$, $\phi_H=0.3$, $r_H=1.0$ can be visualized in Fig.~\ref{plot_Vph} (e).

\begin{table}[H]
\begin{center}
\begin{tabular}{ |c|c|c|c|c|c|c|c|c| } 
 \hline
 $\phi_H$ &  $b_c$ & $b^{-}_2$ & $b^{+}_2$  & $b^{-}_3$  & $b^{+}_3$  \\ 
\hline
 0 (Schwarzschild) & 2.5981  & 2.5005   & 3.0330   & 2.5860  & 2.6050     \\
 \hline
 0.2 & 2.6391  & 2.5490   & 3.0996   &  2.6350 &  2.6546    \\
\hline
 0.4 &  2.8100 & 2.7128   & 3.3219  & 2.8056  & 2.8269     \\
 \hline
 0.45 &  2.9223  & 2.8205   & 3.4652  & 2.9177  & 2.9402      \\
 \hline
 0.4999 & 3.9452   & 3.7951   & 4.7619  & 3.9383  & 3.9716     \\
 \hline
\end{tabular}
\caption{The values of impact parameters associated to different types of the null geodesics around the HBH with $\phi_1=0.5$, $r_H=1.0$. \label{ta3}}
\end{center} 
\end{table}

\section{Conclusion and Outlook}\label{sec:con}

In conclusion, we have studied the geodesic motion of test particles around a spherically symmetric and asymptotically flat hairy black hole (HBH) in the Einstein-Klein-Gordon (EKG) theory. The scalar potential $V(\phi)$ contains asymmetric vacua with the true vacuum being $\phi_1$ and the false vacuum being $\phi=0$. When the scalar field $\phi_H$ is nontrivial at the horizon for $0\leq \phi_H <\phi_1$, the solutions of HBH can exhibit some new properties different from the Schwarzschild black hole, as a consequence from the evasion of the no-hair theorem. In our investigation, we fix $\phi_1=0.5, 1.0$. Since the spacetime of HBH is spherically symmetric, we can obtain the effective potential $V_{\text{eff}}(r)$ of test particles as a function of radial coordinate $r$, and this allows us to focus on the motion of test particles on the equatorial plane. The analysis on the profiles of $V_{\text{eff}}(r)$ could allow us to know the possible orbits of test particles and the integration of the geodesic equation allows us to obtain their trajectories. 

For the case of a massive test particle around the HBHs with $r_H=1$ and $\phi_1=0.5, 1.0$, initially $V_{\text{eff}}(r)$ is a strictly increasing function when the angular momentum squared $L^2$ is very small. However, the increase of $L^2$ can change this monotonically increasing behaviour, until it possesses an inflection point when $L^2$ reaches to a critical value, $L^2_\text{ISCO}$ which is known as the innermost stable circular orbit (ISCO) where the particle can stay in a circular orbit with a minimal radius $r_{\text{ISCO}}$ without being absorbed by the HBH and escaping to infinity. The deviation for the values of $L^2_\text{ISCO}$ and $r_{\text{ISCO}}$ of the HBHs from the Schwarzschild black hole can be increased as $\phi_H$ increases for large $\phi_1$. When $L^2$ exceeds $L^2_\text{ISCO}$, $V_{\text{eff}}(r)$ develops a local maximum and local minimum, indicating the presence of unstable and stable circular orbits, respectively. Besides, the particle can travel in a bound orbit and an escape orbit.  

For the photon around the HBHs with $r_H=1$ and $\phi_1=0.5, 1.0$, its $V_{\text{eff}}(r)$ possesses a local maximum, which indicates the presence of unstable photon sphere. The location of photon sphere around the HBHs doesn't deviate too much from the Schwarzschild black hole, perhaps the increment of $\phi_1$ could exhibit large deviation. According to the total number of orbits, the null geodesics can be divided into three types which are the direct, lensing and photon sphere with their associated impact parameters $b$. When the strength of scalar field at the horizon increases, the values of $b$ correspond to these three types of null geodesics can vary significantly from the Schwarzschild black hole.

Finally, several new research directions can be carried out from this work for the future research. First, we can continue to study the imaging of shadow with different emission profiles for this HBH, and compare them with the Schwarzschild black hole, since we have already calculated the null geodesics in this paper. Recently, the shadow of HBH with inverted Higgs-like scalar potential \cite{Chew:2023olq} has been studied \cite{Lim:2025cne}. Then, we can also study the imaging of shadow and then provide some constraints for some parameters for different HBHs \cite{Chew:2024rin,Chew:2024evh} with different scalar potentials in the EKG theory.

\section*{Acknowledgement}
XYC is supported by the starting grant of Jiangsu University of Science and Technology (JUST).  WF is supported in part by the National Natural Science Foundation of China under Grant No. 12105121. We acknowledge to have useful discussions with Carlos Benavides-Gallego, Jose Luis Bl\'azquez-Salcedo, Jutta Kunz, Eduard Alexis Larranaga.

%\bibliography{mybiblio}

\begin{thebibliography}{99}
\bibitem{schval1}
The value for the location of ISCO of Schwarzcshild black hole in the radial coordinate $r$ is given by $r_\text{ISCO}=3r_H$.
\bibitem{schval2}
The value for the location of photon sphere of Schwarzcshild black hole in the radial coordinate $r$ is given by $r_\text{ph}=3r_H/2$.
\end{thebibliography}

\end{document}